\documentstyle[preprint,aps,epsfig]{revtex} 
\newcommand{\be}[1]{\begin{equation} \label{(#1)}} 
\newcommand{\ee}{\end{equation}}  
\newcommand{\ba}[1]{\begin{eqnarray} \label{(#1)}} 
\newcommand{\ea}{\end{eqnarray}} 
\newcommand{\nn}{\nonumber}

\def\Lfv{$L_f\hspace{-0.95em}/\ \ $}

\def\m{$\mu^--e^-$}

\tightenlines

\begin{document}

\title{
\hfill {\bf Preprint USM-TH-153}  \\[1.5cm]
Effective Lagrangian approach to nuclear $\mu^--e^-$ 
conversion \\and the role of vector mesons.} 

\author{Amand \ Faessler \footnotemark[1], 
Th. \ Gutsche \footnotemark[1], 
Sergey \ Kovalenko \footnotemark[2],\\
V. \ E. \ Lyubovitskij \footnotemark[1]\footnote[3]
{On leave of absence from Department of Physics, 
Tomsk State University, 634050 Tomsk, Russia}, 
Ivan \ Schmidt \footnotemark[2], 
F. \v Simkovic \footnotemark[1]\footnote[4]{On leave of 
absence from Department of Nuclear Physics, Comenius University,
Mlynsk\'a dolina F1, SK--842 15 Bratislava, Slovakia} 
\vspace*{0.4\baselineskip}}
\address{
\footnotemark[1] 
Institut f\"ur Theoretische Physik, Universit\"at T\"ubingen, \\
Auf der Morgenstelle 14, D-72076 T\"ubingen, Germany 
\vspace*{0.2\baselineskip}\\
\footnotemark[2]
Departamento de F\'\i sica, Universidad
T\'ecnica Federico Santa Mar\'\i a, \\ 
Casilla 110-V, Valpara\'\i so, Chile
\vspace*{0.3\baselineskip}\\}

\date{\today}
 
\maketitle 
 
\vskip.5cm 

\begin{abstract}
We study nuclear $\mu^--e^-$ conversion in the general framework of an 
effective Lagrangian approach without referring to any specific 
realization of the physics beyond the standard model (SM) responsible 
for lepton flavor violation (\Lfv). 
We examine the impact of a specific hadronization prescription on 
the analysis of new physics in nuclear $\mu^--e^-$ conversion and stress
the importance of vector meson exchange between lepton and nucleon 
currents. A new issue of this mechanism is the presence of the strange 
quark vector current contribution induced by the $\phi$ meson.
This allows us to extract new limits on the \Lfv lepton-quark 
effective couplings from the existing experimental data.
\end{abstract}

\vskip .3cm
 


\section{Introduction}

The discovery of neutrino oscillations has established the fact that 
Nature does not respect lepton flavor conservation contrary to the 
expectations within the Standard Model (SM). However, this is so far 
the only experimental observation of the violation of this conservation 
law. 
On the other hand, once \Lfv is proved, it is natural to expect its 
manifestations in the charged lepton sector as well. The smallness 
of neutrino mass square differences, $\Delta m^2$, observed in neutrino 
oscillation experiments makes the neutrino induced \Lfv effects 
(tree-level exchange, loops, boxes) in the processes with charged leptons 
extremely suppressed as $(\Delta m^2)^2/M_W^4$, pushing them far beyond 
the reach of experimental searches\footnote{This conclusion is valid for 
the conventional three light neutrino mixing scenarios. The incorporation 
of sterile neutrinos may significantly reduce the suppression 
factor~\cite{Simkovic:2001fy}.}. However, the charged lepton processes may 
receive another \Lfv contributions from the physics beyond the SM 
attributed to a certain high-energy \Lfv scale $\Lambda_{LFV}$.
Thus, searching for the lepton flavor violation in reactions with 
charged leptons becomes challenging both from the experimental and 
theoretical points of view. 

Muon-to-electron ($\mu^--e^-$) conversion in nuclei 
\begin{equation} 
\mu^- + (A,Z) \longrightarrow  e^- \,+\,(A,Z)^\ast 
\label{I.1} 
\end{equation} 
is commonly recognized as one of the most promising probes of lepton 
flavor violation in the charged lepton sector and of related physics 
beyond the SM (for reviews, 
see~\cite{Kosmas:ch,Marciano:conf,Kuno:1999jp}). 
This is, in particular, due to the very high sensitivity of the 
experiments dedicated to search for this process. 

At present there is one running experiment, 
SINDRUM II~\cite{Honecker:zf}, 
and two planned ones, MECO~\cite{Molzon:sf,MECO} and PRIME~\cite{PRIME}. 
The SINDRUM II experiment at PSI~\cite{Honecker:zf} with $^{48}$Ti 
as stopping target has established the best upper bound on the 
branching ratio~\cite{Honecker:zf}  
\begin{eqnarray}\label{Ti} 
&&R_{\mu e}^{Ti} = \frac{\Gamma(\mu^- + {}^{48}Ti\rightarrow e^- + 
{}^{48}Ti)} {\Gamma(\mu^- + {}^{48}Ti 
\rightarrow \nu_{\mu} + ^{48}Sc)} \leq 6.1\times 10^{-13}\ , \ \ \ 
\mbox{(90\% C.L.)} \,.  
\end{eqnarray}
The MECO experiment with $^{27}$Al is going to start soon at 
Brookhaven~\cite{MECO}. The sensitivity of this experiment is expected 
to reach the level of 
\begin{eqnarray}\label{Al}
R_{\mu e}^{Al}  = \frac{\Gamma(\mu^- + {}^{27}Al\rightarrow e^- 
+  {}^{27}Al)}{\Gamma(\mu^- + {}^{27}Al
\rightarrow\nu_{\mu} + {}^{27}Mg)}\leq 2\times 10^{-17}\ \ \ \ \ \ \ \  
 \end{eqnarray}
The PSI experiment is also running with the very heavy nucleus 
$^{197}$Au aiming to improve the previous limit by the same 
experiment~\cite{Honecker:zf,Vintz} up to 
\begin{eqnarray}\label{Au} 
R_{\mu e}^{Au} = \frac{\Gamma(\mu^- + ^{197}Au\rightarrow e^- 
+ ^{197}Au)}{\Gamma(\mu^- + ^{197}Au \rightarrow 
\nu_{\mu} + ^{197}Pt)} \leq 6 \times 10^{-13}\ \ \ \ \ \ \  
\end{eqnarray}
The proposed new experiment PRIME (Tokyo)~\cite{PRIME} is going to 
utilize $^{48}$Ti as stopping target with an expected sensitivity 
of 
\begin{eqnarray}\label{Prime} 
R_{\mu e}^{Ti} = \frac{\Gamma(\mu^- + {}^{48}Ti\rightarrow e^- + 
{}^{48}Ti)} {\Gamma(\mu^- + {}^{48}Ti 
\rightarrow \nu_{\mu} + ^{48}Sc)} \leq  10^{-18}. 
\end{eqnarray}
These experimental bounds would allow to set stringent limits on 
the mechanisms of \m conversion and the underlying theories of \Lfv. 
In the literature various mechanisms beyond the SM have been studied 
(see~\cite{Kosmas:ch,Marciano:conf,Kuno:1999jp} and references therein). 
They can be classified as photonic and non-photonic, that is with and 
without photon exchange between the lepton and nuclear vertices, 
respectively. These two categories of mechanisms differ significantly 
from each other in various respects. In fact, they receive  
different contributions from the new physics and also require different 
treatments of the effects of the nucleon and the nuclear structure. 
Latter aspect is, in particular, attributed to the fact that the two 
mechanisms operate at different distances and, therefore, involve 
different details of the nucleon and nuclear structure. 

Long-distance photonic mechanisms (Fig.1a) are mediated by virtual
photon exchange between the nucleus and the $\mu-e$ lepton current.
They suggest that the \m conversion occurs in the lepton-flavor 
non-diagonal electromagnetic vertex which is presumably induced by 
non-standard model physics at the loop level. The hadronic vertex is 
characterized in this case by ordinary electromagnetic nuclear 
form factors. Contributions to \m conversion via virtual photon exchange 
exist in all models which allow $\mu\rightarrow e \gamma$ decay. 
On the other hand, short-distance non-photonic mechanisms (Fig.1b) are 
described by the effective \Lfv $ $ 4-fermion quark-lepton interactions 
which may appear after integrating out heavy intermediate states ($W,Z$, 
Higgs bosons, supersymmetric particles etc.). 

In this paper we concentrate on the non-photonic mechanisms of \m \, 
conversion. The generic \Lfv effects of physics beyond the SM are 
parameterized by an effective Lagrangian with all possible 4-fermion 
quark-lepton interactions consistent with Lorentz covariance and gauge 
symmetry. We pay special attention to the hadronization of the quark 
currents of this Lagrangian  focusing on its special realization when 
quarks are embedded into meson fields. 
Previously, in Ref.~\cite{Faessler:2004jt}, we have shown that this 
realization is especially relevant for vector interactions which 
receive an appreciable contribution from vector meson exchange. 
This result uncovers the important role of vector mesons in the analysis 
of new physics in \m conversion. Below we present a detailed discussion 
of the vector meson exchange mechanism of \m nuclear conversion and 
examine some basic assumptions underlying the choice of the effective 
Lagrangian of \Lfv meson-lepton interactions.  

\section{General Framework}

We start with the 4-fermion effective Lagrangian describing the 
non-photonic $\mu^- - e^-$ conversion at the quark level. The most 
general form of this Lagrangian has been derived in 
Ref.~\cite{Kosmas:2001mv}. Here we present only those terms which 
contribute to the coherent $\mu^- - e^-$ conversion:
\begin{eqnarray} 
{\cal L}_{eff}^{lq}\ =\  \frac{1}{\Lambda_{LFV}^2}  
\left[(\eta_{VV}^{q} j_{\mu}^V\ + \eta_{AV}^{q} 
j_{\mu}^A) J_{q}^{V\mu} + 
(\eta_{SS}^{q} j^S\ + \eta_{PS}^{q} j^P) J_{q}^{S}\right], 
\label{eff-q}
\end{eqnarray}
where $\Lambda_{LFV}$ is the characteristic high energy scale of 
lepton flavor violation attributed to new physics. The summation runs 
over all the quark species $q= \{u,d,s,c,b,t\}$.  
Lepton and quark currents are defined as:  
\begin{eqnarray}\label{currents-1}
j_{\mu}^V = \bar e \gamma_{\mu} \mu\,, \,\,\,\,\,  
j_{\mu}^A = \bar e \gamma_{\mu} \gamma_{5} \mu\,, \,\,\,\,\,  
j^S = \bar e \ \mu\,, \,\,\,\,\, 
j^P = \bar e \gamma_{5} \mu\,, \,\,\,\,\,
J_{q}^{V\mu} = \bar q \gamma^{\mu} q\,, \,\,\,\,\,
J_{q}^{S} = \bar q \ q \,.
\end{eqnarray} 
The \Lfv parameters $\eta^{q}$ in Eq.~(\ref{eff-q}) 
depend on a concrete \Lfv model. 

The next step is the derivation of a Lagrangian in terms of effective 
nucleon fields which is equivalent to the quark level 
Lagrangian~(\ref{eff-q}). First, we write down the lepton-nucleon \Lfv 
Lagrangian of the coherent $\mu^- - e^-$ conversion in a general 
Lorentz covariant form with the isospin structure of the \m transition 
operator~\cite{Kosmas:2001mv}: 
\begin{eqnarray} 
\hspace*{-.5cm} 
{\cal L}_{eff}^{lN} =   \frac{1}{\Lambda_{LFV}^2} 
\left[j_{\mu}^a (\alpha_{aV}^{(0)} J^{V\mu \, (0)} + 
\alpha_{aV}^{(3)} J^{V\mu \, (3)}) + j^b (\alpha_{bS}^{(0)} 
J^{S \, (0)} + \alpha_{bS}^{(3)} J^{S \, (3)})\right]\,, 
\label{eff-N} 
\end{eqnarray} 
where the summation runs over the double indices $a = V,A$ and 
$b = S,P$. The isoscalar $J^{(0)}$ and isovector $J^{(3)}$ 
nucleon currents are defined as 
\begin{eqnarray}\label{currents-2}
J^{V\mu \, (k)} \, = \, \bar N \, \gamma^\mu \, \tau^k \,  N, \ \ \  
J^{S \, (k)}  \, = \, \bar N \, \tau^k \, N\,,
\end{eqnarray} 
where $N$ is the nucleon isospin doublet, $ k = 0,3 $ and 
$\tau_0\equiv\hat I$. 

This Lagrangian is supposed to be generated by the one of 
Eq.~(\ref{eff-q}) and, therefore, must correspond to the same  
order $1/\Lambda_{LFV}^{2}$ in inverse powers of the \Lfv scale. 
The Lagrangian~(\ref{eff-N}) is the basis 
for the derivation of the nuclear transition operators. 

Now one needs to relate the lepton-nucleon \Lfv parameters $\alpha$  
in Eq.~(\ref{eff-N}) to the more fundamental lepton-quark \Lfv 
parameters $\eta$ in Eq.~(\ref{eff-q}). This implies a certain 
hadronization prescription which specifies the way in which the effect 
of quarks is simulated by hadrons. In the absence of a true theory of 
hadronization we rely on some reasonable assumptions and models. 

There are basically two possibilities for the hadronization mechanism. 
The first one is a direct embedding of the quark currents into the 
nucleon (Fig.2a), which we call direct nucleon mechanism (DNM).  
The second possibility is a two stages process (Fig.2b). First, the   
quark currents are embedded into the interpolating meson fields which
then interact with the nucleon currents. We call this possibility 
meson-exchange mechanism (MEM). 
In general one expects all the mechanisms to contribute 
to the coupling constants $\alpha$ in Eq.~(\ref{eff-N}). However, 
at present the relative amplitudes of each mechanism are unknown. 
In view of this problem one may try to understand the importance of 
a specific mechanism, assuming for simplicity, that only this 
mechanism is operative and estimating its contribution to the process 
in question. We follow this procedure for the case of $\mu^- - e^-$ 
conversion and consider separately the contributions of the direct 
nucleon mechanism $\alpha_{[N]}$ and the meson-exchange one 
$\alpha_{[MN]}$ to the couplings of the Lagrangian (\ref{eff-N}).  

In the present paper we concentrate on the meson-exchange mechanism and 
compare its contribution to \m conversion with that of the direct 
nucleon mechanism which we shortly review in the next section following 
Ref.~\cite{Kosmas:2001mv}.

\section{Direct Nucleon Mechanism}

As mentioned before, this mechanism relies on a direct embedding of 
the quark currents of the Lagrangian (\ref{eff-q}) into the 
corresponding nucleon currents. This hadronization prescription directly 
leads to the nucleon level Lagrangian in Eq.~(\ref{eff-N}) describing 
a contact type $\mu^--e^-$ conversion as shown in Fig.2a. 

Now we relate the coefficients $\alpha$ in Eq.~(\ref{eff-N}) with the 
"fundamental" \Lfv parameters $\eta$ of the quark level 
Lagrangian (\ref{eff-q}). Towards this end we apply the on-mass-shell 
matching condition~\cite{Faessler:1996ph}
\begin{equation}
\langle e^-N|{\cal L}_{eff}^{q}|\mu^-N\rangle \approx
 \langle e^-N|{\cal L}_{eff}^{N}|\mu^-N\rangle ,  
\label{match}
\end{equation}
in terms of the matrix elements of the Lagrangians (\ref{eff-q}) and 
(\ref{eff-N}) between the initial and final states of \m conversion at 
the nucleon level.

In order to solve this equation in Ref.~\cite{Kosmas:2001mv} various 
relations for the matrix elements of quark operators between nucleon 
states were used 
\begin{eqnarray}\label{mat-el1}
\langle N|\bar{q}\ \Gamma_{K}\ q|N\rangle = G_{K}^{(q,N)}
\bar{N}\ \Gamma_{K}\ N,
\end{eqnarray}
with $q=\{u,d,s\}$,  $N=\{p,n\}$. 
Since the maximum momentum transfer in $\mu -e$ conversion is much 
smaller than the typical scale of nucleon structure one can safely 
neglect the ${\bf q}^{2}$-dependence of the nucleon form factors 
$G_{K}^{(q,N)}$.

Isospin symmetry requires that
\begin{eqnarray}\label{isosym}
&&
G_{K}^{(u,p)}=G_{K}^{(d,n)}\equiv G_{K}^{u}, \ \ \
G_{K}^{(d,p)}=G_{K}^{(u,n)}\equiv G_{K}^{d}, \\ \nn
&&G_{K}^{(s,p)}=G_{K}^{(s,n)}\equiv G_{K}^{s}, \ \ \ 
G_{K}^{(h,p)}=G_{K}^{(h,n)}\equiv G_{K}^{h}.
\end{eqnarray}                    
Here $h = c,b,t$ are the heavy quarks.

The vector quark currents in the non-relativistic limit result in the 
quark number operators and, thus, the vector form factors at $q^2=0$ 
are equal to the total number of the corresponding species of quarks 
in the nucleon. Therefore, 
\begin{eqnarray}\label{V-FFN}
G_{V}^{u}=2,\ \ \  G_{V}^{d}=1, \ \ \ G_{V}^{s}=0, \ \ \ G_V^{h}=0.
\end{eqnarray}
Because of the last two equalities $s, c, b, t$ quarks 
do not contribute to the couplings of the vector 
nucleon current in Eq.~(\ref{eff-N}). 

Now solving Eq.~(\ref{match}) with the help of 
Eqs.~(\ref{mat-el1})-(\ref{V-FFN}) one can express the coefficients 
$\alpha$ of the nucleon level Lagrangian (\ref{eff-N}) in terms of the 
generic \Lfv parameters $\eta$ of the quark level effective Lagrangian 
Eq.~(\ref{eff-q}). Here we present only the results of 
Ref.~\cite{Kosmas:2001mv} relevant for our analysis which are the 
couplings of the vector nucleon currents in Eq.~(\ref{eff-N}): 
\begin{eqnarray} 
\label{alpha}
\hspace*{-.65cm}
\alpha_{aV[N]}^{(3)} = \frac{1}{2}(\eta_{aV}^{u} - \eta_{aV}^{d}) 
(G_{V}^{u} - G_{V}^{d})\,, 
\,\,\,  
\alpha_{aV[N]}^{(0)}  = \frac{1}{2}(\eta_{aV}^{u} + \eta_{aV}^{d}) 
(G_{V}^{u} + G_{V}^{d})\,,
\end{eqnarray} 
where $a=V,A$. 

Concluding this section we stress the fact that $s, c, b, t$ quarks 
do not contribute to the couplings of the vector nucleon current in 
Eq.~(\ref{alpha}). In the next section we will show that the vector 
meson exchange may drastically modify this situation and introduce 
the contribution of strange quarks into these couplings.

\section{Vector Meson Contribution} 

Now let us turn to the contributions of the meson-exchange mechanism to 
the couplings $\alpha$ of the lepton-nucleon Lagrangian (\ref{eff-N}).  
The mesons which can contribute to this mechanism are the unflavored 
vector and scalar ones. Since the case of the scalar meson candidate 
$f_0(600)$ is still quite uncertain~\cite{Hagiwara:fs} we do not study 
its contribution. Thus, we are left with the vector mesons. The lightest 
of them, giving the dominant contributions, are the isovector 
$\rho(770)$ and the two isoscalar $\omega(782), \ \phi(1020)$ mesons. 
In our analysis we adopt ideal singlet-octet mixing corresponding to 
the following quark content of the $\omega$ and $\phi$ 
mesons~\cite{Hagiwara:fs}:  
\begin{eqnarray}\label{quark-cont}
\omega = (u \bar u + d \bar d)/\sqrt{2}, \ \ \ \phi = - s \bar s\,.
\end{eqnarray} 
We estimate the vector meson contribution to the lepton-nucleon 
Lagrangian (\ref{eff-N}) in two ways. First we adopt a model 
independent effective Lagrangian approach and then we present a more 
restrictive approach based on a simplified model of hadronization. 
The latter case results in a significant suppression of the vector 
meson contribution. A comparison of 
both approaches allows us to get an idea of the impact of the 
hadronization procedure on the new physics contribution to 
\mbox{$\mu^- - e^-$}conversion and the reliability of the limits on 
the corresponding parameters derived from the experimental data.

\subsection{Model-Independent Approach}

First, we derive the \Lfv lepton-meson effective Lagrangian in terms 
of the interpolating $\rho^0, \ \omega$ and $\phi$ fields retaining 
all the interactions consistent with Lorentz and electromagnetic 
gauge invariance. It can be written as: 
\begin{eqnarray}\label{eff-LV}
{\cal L}_{eff}^{lV}\ &=& \  \frac{\Lambda_H^2}{\Lambda_{LFV}^2}  
\biggl[ \, \biggl\{ (\xi_V^{\rho} j_{\mu}^V\ + \xi_A^{\rho} 
j_{\mu}^A\ )\rho^{0\ \mu} + (\xi_V^{\omega}  j_{\mu}^V\ 
+ \xi_A^{\omega} j_{\mu}^A\ )\omega^{\mu} + \\ \nn 
 &+& (\xi_V^{\phi}  j_{\mu}^V\ + \xi_A^{\phi} 
j_{\mu}^A\ )\phi^{\mu} \biggr\}  + \frac{1}{\Lambda_H^2}
\biggl\{ \xi_V^{\rho(2)} j_{\mu}^V\ 
\partial^{\mu}\partial^{\nu} \rho^0_{\nu} + ... \biggr\} + ... \biggr], 
\end{eqnarray}  
with the unknown dimensionless coefficients $\xi$ to be determined 
from the had\-ro\-ni\-za\-ti\-on prescription. Since this Lagrangian is 
supposed to be generated by the quark-lepton Lagrangian (\ref{eff-q})  
all its terms have the same suppression $\Lambda_{LFV}^{-2}$ with 
respect to the large \Lfv scale $\Lambda_{LFV}$. Another scale in the 
problem is the hadronic scale $\Lambda_H \sim 1$ GeV which adjusts the 
physical dimensions of the terms in  Eq.~(\ref{eff-LV}). Typical 
momenta involved in $\mu^- - e^-$ conversion are $q \sim m_{\mu}$, 
where $m_{\mu}$ is the muon mass. Thus, from naive dimensional counting 
one expects that the contribution of the derivative terms to 
$\mu^- - e^-$ conversion is suppressed by a factor 
$(m_{\mu}/\Lambda_H)^2\sim 10^{-2}$ in comparison to the contribution 
of the non-derivative terms. Therefore, at this step in 
Eq.~(\ref{eff-LV}) we retain only the dominant non-derivative terms. 
However, it is worth noting that a true hadronization theory, 
yet non-existing, may forbid such terms so that the expansion in 
Eq.~(\ref{eff-LV}) starts from the derivative terms. 
Later on we demonstrate how it happens in a particular model.  

In order to relate the parameters $\xi$ of the Lagrangian (\ref{eff-LV}) 
with the ``fundamental" parameters $\eta$ of the quark-lepton Lagrangian 
(\ref{eff-q}) we use, as in the previous section, an approximate method 
based on the standard on-mass-shell matching 
condition~\cite{Faessler:1996ph} which in this case takes the form  
\begin{equation}\label{match1} 
\langle \mu^+ \, e^-|{\cal L}_{eff}^{lq}|V\rangle \approx 
\langle \mu^+ \, e^-|{\cal L}_{eff}^{lV}|V \rangle ,  
\end{equation} 
with $|V= \rho, \omega, \phi \rangle$ corresponding to vector meson 
states on their mass-shell. We solve equation (\ref{match1}) using the 
well-known quark current matrix elements 
\begin{eqnarray}\label{mat-el2} 
&&\langle 0|\bar u \, \gamma_\mu \, u|\rho^0(p,\epsilon)\rangle \, = \, 
\, - \, \langle 0|\bar d \, \gamma_\mu \, d|\rho^0(p,\epsilon)\rangle 
\, = \, m_{\rho}^2 \, f_{\rho} \, \epsilon_\mu(p), \nonumber\\ 
&&\langle 0|\bar{u}\ \gamma_{\mu}\ u|\omega(p,\epsilon)\rangle = 
\langle 0|\bar{d}\ \gamma_{\mu}\ d|\omega(p,\epsilon)\rangle 
\, = \, 3 \, m_{\omega}^2 \, f_{\omega} \, \epsilon_\mu(p)\,,\\ 
&&\langle 0|\bar{s}\ \gamma_{\mu}\ s|\phi(p,\epsilon)\rangle 
\, = \, - \, 3 \, m_\phi^2 \, f_{\phi} \, \epsilon_\mu(p) \,.  
\nonumber
\end{eqnarray} 
Here $p$, $m_V$ and $\epsilon_{\mu}$ are the vector-meson four-momentum, 
mass and the polarization state vector, respectively. The quark 
operators in Eq.~(\ref{mat-el2}) are taken at $x=0$. The coupling 
constants $f_{V}$ are determined from the $V\to e^+ e^-$ decay width:  
\begin{eqnarray}\label{decay}
\Gamma(V \to e^+ e^-) \, = \, \frac{4 \pi}{3} \, \alpha^2 \, 
f_{V}^2 \, m_V, 
\end{eqnarray} 
where $\alpha$ is the fine-structure constant.
The current central values of the meson couplings $f_{V}$ 
and masses $m_V$ are~\cite{Hagiwara:fs}: 
\begin{eqnarray}\label{constants}  
&&\hspace*{.15cm}f_{\rho}   = 0.2 \,, \hspace*{1.7cm} 
  f_{\omega} = 0.059  \,, \hspace*{1.5cm} 
  f_{\phi}   = 0.074  \,\,, \\   
&&m_{\rho}   = 771.1 \,\,  \mbox{MeV}, \,\,\,
  m_{\omega} = 782.57 \,\, \mbox{MeV}, \,\,\,  
  m_{\phi}   = 1019.456 \,\, \mbox{MeV}\,. \nonumber
\end{eqnarray} 
Solving Eq.~(\ref{match1}) with the help of Eqs.~(\ref{mat-el2}), 
we obtain the desired expressions for the coefficients $\xi$ of the 
lepton-meson Lagrangian (\ref{eff-LV}) in terms of generic \Lfv  
parameters $\eta$ of the lepton-quark effective Lagrangian 
Eq.~(\ref{eff-q}): 
\begin{eqnarray}
&&\xi_a^{\rho} \, = \, \left(\frac{m_{\rho}}{\Lambda_H}\right)^2 
f_{\rho} \,(\eta_{a V}^{u} - \eta_{a V}^{d})\,, \,\, 
\xi_a^{\omega} \, = \, 3 \left(\frac{m_{\omega}}{\Lambda_H}\right)^2  
\, f_{\omega} \, (\eta_{a V}^{u} + \eta_{a V}^{d}) \,, \,\,\\ \nn
&&\xi_a^{\phi} \, = \, - 3 \left(\frac{m_{\phi}}{\Lambda_H}\right)^2  
\, f_{\phi} \, \eta_{a V}^{s} \, ,
\end{eqnarray} 
where $a = V, A$. 

Now we derive the vector meson exchange contributions 
to the couplings of the effective Lagrangian in 
Eq.~(\ref{eff-N}) expressed in terms of nucleon fields.
To this end we introduce the effective Lagrangian 
describing the interaction of nucleons with vector 
mesons~\cite{Weinberg:de,Mergell:1995bf,Kubis:2000zd}: 
\begin{eqnarray}\label{MN} 
{\cal L}_{VN} \, = \, \frac{1}{2} \, 
\bar{N}\gamma^{\mu}\left[ g_{_{\rho NN}} \, \vec{\rho}_{\mu} \, 
\vec{\tau} \,  + \, g_{_{\omega NN}} \, \omega_{\mu} \, + \, 
g_{_{\phi NN}} \, 
\phi_{\mu}\right] N\,. 
\end{eqnarray} 
In this Lagrangian we neglected the derivative terms, irrelevant for 
coherent $\mu^- - e^-$ conversion. For the meson-nucleon couplings 
$g_{VNN}$ we use numerical values taken from an updated dispersive 
analysis~\cite{Mergell:1995bf,Meissner:1997qt} 
\begin{eqnarray}\label{VN-couplings}
g_{_{\rho NN}}= 4.0\,, \,\, 
g_{_{\omega NN}} = 41.8\,, \,\, 
 , \ \ \    
g_{_{\phi NN}}^{(1)}= - 18.3, \ \ \ 
g_{_{\phi NN}}^{(2)} = - 0.24\,. 
\end{eqnarray} 
At this point the following comment is in order. The relatively large 
value of the ${\phi NN}$  coupling $g_{_{\phi NN}}^{(1)} = - 18.3$ in 
Eq.~(\ref{VN-couplings}) has been derived in Ref.~\cite{Mergell:1995bf} 
on the basis of the assumption on the "maximal" violation of the 
Okuba-Zweig-Iizuka (OZI) rule. It was also stressed in 
Ref.~\cite{Mergell:1995bf} that this value corresponds to the upper 
limit for the ${\phi NN}$ coupling which parameterizes the full spectral 
function in the mass region of $\sim$1 GeV within the $\phi$ pole 
dominance approximation. The inclusion of other contributions such as 
the $\pi\rho$ continuum leads to a significant reduction of the 
$g_{_{\phi NN}}$ coupling~\cite{Meissner:1997qt}. The detailed 
analysis of various meson and baryon cloud contributions to the vector 
${\phi NN}$ coupling results in the value 
$g_{_{\phi NN}}^{(2)} = - 0.24$~\cite{Meissner:1997qt}. 
For completeness we consider both values of $g_{_{\phi NN}}$ coupling  
presented in Eq.~(\ref{VN-couplings}) in our numerical analysis.   
Regarding these two values of the $g_{_{\phi NN}}$ coupling the 
following comments are in order. The "maximal" value of the 
$g_{_{\phi NN}}$ coupling is compatible with SU(3) symmetry prediction. 
Substituting the values of the $g_{\rho NN}$, $g_{\omega NN}$ 
and $g_{_{\phi NN}}^{(1)}$ constants from 
Eq.~(\ref{VN-couplings}) into the SU(3) relation~\cite{deSwart:gc} 
\begin{eqnarray}\label{SU3}
g_{_{\phi NN}} \, = \, g_{_{\rho NN}} \, (\sqrt{3}/\cos\theta_V) \, 
- \, g_{_{\omega NN}} \, \tan\theta_V,
\end{eqnarray}
we estimate the $\omega-\phi$ mixing angle to be 
$\theta_V = 32.4^{\rm o}$. This value is quite close to the ideal angle  
$\theta_V^I = 35.3^{\rm o}$ and to the experimental one 
$\theta_V^{\rm exp} \simeq 39^{\rm o}$, and therefore, is consistent 
with our initial assumption on the quark content of the $\omega$ and 
$\phi$ mesons (\ref{quark-cont}). It is instructive to estimate the 
SU(3) symmetry prediction for the coupling $g_{_{\phi NN}}$ for the 
angles $\theta_V^I$ and $\theta_V^{\rm exp}$. Substituting
their values to Eq. (\ref{SU3}) with $g_{_{\rho NN}}= 4.0$ and 
$g_{_{\omega NN}} = 41.8$ one finds the corresponding values of the 
$\phi$-nucleon coupling 
\begin{eqnarray}\label{IE}
\theta_V^I:\ \ \ g_{_{\phi NN}} = -21.1,\ \ \ \ \theta_V^{\rm exp}: 
\ \ \ g_{_{\phi NN}} = -24.9,
\end{eqnarray}
which are larger than $g_{_{\phi NN}}^{(1)}= - 18.3$ in 
Eq.~(\ref{VN-couplings}).

On the other hand the value $g_{_{\phi NN}}^{(2)} = - 0.24 $ results in 
a strong violation of SU(3) symmetry and Eq.~(\ref{SU3}) is no longer 
valid giving a very small estimate for $\theta_V \simeq 9.9^{\rm o}$. 
This result looks controversial in view of successful predictions of 
the approximate SU(3) symmetry for the masses of $\omega$ and $\phi$ 
mesons with the mixing angle close to its ideal value. We assume that 
the true value of $g_{_{\phi NN}}^{\rm}$ lies in the interval 
$-18.3 \leq g_{_{\phi NN}}^{\rm} \leq -0.24$.

The vector meson-exchange contribution to the nucleon-lepton effective 
Lagrangian (\ref{eff-N}) arises in second order in the Lagrangian 
${\cal L}_{eff}^{lV} + {\cal L}_{VN}$ and corresponds to the diagram 
in Fig.2b. We estimate this contribution only for the coherent 
$\mu^- - e^-$ conversion process. In this case we disregard all the 
derivative terms of nucleon and lepton fields. Neglecting the kinetic 
energy of the final nucleus, the muon binding energy and the electron 
mass, the square momentum transfer $q^2$ to the nucleus has a constant 
value $q^2 \approx - m_{\mu}^2$. In this approximation the vector meson 
propagators convert to $\delta$-functions leading to effective 
lepton-nucleon contact type operators. Comparing them with the 
corresponding terms in the Lagrangian (\ref{eff-N}), we obtain for the 
vector meson-exchange contribution to the coupling constants:
\begin{eqnarray} \label{alpha-V-ex}
\alpha_{aV[MN]}^{(3)} &=& - \beta_{\rho}(\eta_{aV}^{u} - \eta_{aV}^{d}),
\ \ \  
\alpha_{aV[MN]}^{(0)} = - \beta_{\omega}(\eta_{aV}^{u} + \eta_{aV}^{d}) 
- \beta_{\phi}\eta_{aV}^{s}\,,
\end{eqnarray}
with $a=V,A$ and the coefficients 
\begin{eqnarray} \label{beta}
\beta{_{\rho}} = \frac{1}{2} 
\frac{g_{_{\rho NN}} \, f_{\rho} \, m_{\rho}^2}{m_{\rho}^2 
+ m_{\mu}^2},\ \beta{_{\omega}} = \frac{3}{2} 
\frac{g_{_{\omega NN}} \, f_{\omega} \, m_{\omega}^2 }
{m_{\omega}^2 + m_{\mu}^2},\ \beta{_\phi} = -\frac{3}{2} 
\frac{g_{_{\phi NN}} \, f_{\phi} \, 
m_{\phi}^2 }{m_{\phi}^2 + m_{\mu}^2}. 
\end{eqnarray}
Substituting the values of the meson coupling constants and masses 
from Eqs.~(\ref{constants}) and (\ref{VN-couplings}), and including 
the two different options for the $g_{_{\phi NN}}$ constant, 
we obtain for these coefficients 
\begin{eqnarray} \label{beta-num} 
\beta{_{\rho}} = 0.39\,, \,\,\, 
\beta{_{\omega}} = 3.63\,, \,\,\, 
\beta{_{\phi}}^{(1)} = 2.0
\,, \,\,\, 
\beta{_{\phi}}^{(2)} = 0.03 \,\,. 
\end{eqnarray} 
A new issue of the vector meson contribution (\ref{alpha-V-ex})
is the presence of the strange quark vector current contribution 
associated with the LFV parameter $\eta_{aV}^{s}$, absent in the 
direct nucleon mechanism as it follows from Eq.~(\ref{alpha}). 
This opens up the possibility of deriving new limits on this parameter 
from the experimental data on $\mu^- - e^-$ conversion. Another 
surprising result is that the contribution~(\ref{alpha-V-ex}) of 
the meson-exchange mechanism is comparable to the 
contribution~(\ref{alpha}) of the direct nucleon mechanism.

\subsection{Simplified Model of Hadronization} 

Now, instead of constructing the \Lfv lepton-meson Lagrangian in a 
general phenomenological way, only requiring its consistency with basic 
symmetries, as done in the previous subsection, we present a simple 
model which allows us to derive this Lagrangian within certain 
assumptions. The model assumes that the vector mesons fields manifest 
themselves in the interactions with leptons only indirectly via their 
interactions with quarks. Thus, the model Lagrangian consists of two 
terms
\begin{eqnarray} \label{LqV}
{\cal L}_{eff}^{qVl}\, = \,{\cal L}_{eff}^{lq} \,+ \,{\cal L}^{qV}
\end{eqnarray}
where the first term is the \Lfv lepton-quark Lagrangian (\ref{eff-q}) 
and the second one represents the vector meson-quark interaction 
Lagrangian which we write down in a general isospin invariant form as
\begin{eqnarray}\label{V-q}
{\cal L}^{qV} \,= \, \frac{g_{\rho qq}}{2} \,\bar q \, 
\gamma^\mu \, \vec{\tau} \, q \, \vec{\rho}_{\mu} + 
\frac{g_{\omega qq}}{2} \,\bar q \, \gamma^\mu \, q \, \omega_{\mu} +
\frac{g_{\phi qq}}{2} \, \bar s \, \gamma^\mu  s \, \phi_{\mu}.
\end{eqnarray}
Here $q$ is the quark isodoublet. The vector meson-quark couplings 
$g_{_{V qq}}$ are free phenomenological parameters which, however, do 
not appear explicitly in the final results.

The Lagrangian (\ref{LqV}) generates the \Lfv lepton-meson effective 
Lagrangian in second order of perturbation theory via the vector-vector 
quark loops in Fig.3. All the quark loops (QL) are logarithmically 
divergent and transversal. Their Lorentz structure can be written in 
momentum space as  
\begin{eqnarray}\label{loop-1}
\Pi^{\rm QL}_{\mu\nu}(q) \sim (g_{\mu\nu}q^2 - q_{\mu}q_{\nu})\,.
\end{eqnarray} 
This property immediately follows from the observation that these loops 
are nothing but the quark contribution to the vacuum polarization 
operator of the photon, which is transversal. In coordinate space this 
structure is reproduced at the first order in the effective lepton-meson 
Lagrangian with the transversal double derivative operator $(g_{\mu\nu} 
\partial^2 - \partial_{\mu}\partial_{\nu})$ 
acting on meson and/or lepton fields. 
We write down this Lagrangian in an equivalent but more 
conventional form\footnote{There are other equivalent forms of this  
Lagrangian suitable for description of chiral invariant interactions  
of vector mesons with pseudoscalar mesons and the electromagnetic  
field~\cite{Kubis:2000zd,Gasser:1983yg}.}  
\begin{eqnarray}\label{LV-loop}  
{\cal L}_{eff}^{lV} &=&  - \frac{1}{2 \, {\Lambda_{LFV}^2}}  
\biggl[({\tilde\xi}_V^{\rho} j_{\mu\nu}^V + {\tilde\xi}_A^{\rho} 
j_{\mu\nu}^A ) \, {\cal R}^{0 \, \mu\nu} + ({\tilde\xi}_V^{\omega}  
j_{\mu\nu}^V + {\tilde\xi}_A^{\omega}j_{\mu\nu}^A )\,\Omega^{\mu\nu} +\\ 
&+& ({\tilde\xi}_V^{\phi}  j_{\mu\nu}^V + {\tilde\xi}_A^{\phi} 
j_{\mu\nu}^A ) \, \Phi^{\mu\nu} \biggr] \nonumber 
\end{eqnarray}  
in terms of the vector meson and lepton current stress 
tensors~\cite{Kubis:2000zd}: 
\begin{eqnarray}\label{StressT}\nn
&&{\cal R}^{0 \, \mu\nu}= \partial^\mu \rho^{0 \, \nu} - \partial^\nu 
\rho^{0 \, \mu}, \ \  \Omega^{\mu\nu}= \partial^\mu \omega^\nu - 
\partial^\nu \omega^\mu,\ \   
\Phi^{\mu\nu}= \partial^\mu \phi^\nu - \partial^\nu \phi^\mu\,, \\  
&&j_{\mu\nu}^V =  \partial_\mu j_{\nu}^V - \partial_\nu j_{\mu}^V, 
\ \ \ \ \ \ \ 
j_{\mu\nu}^A =  \partial_\mu j_{\nu}^A - \partial_\nu j_{\mu}^A\,.
\end{eqnarray} 
Thus, this model does not allow non-derivative terms in the lepton-meson 
Lagrangian. The relations between the couplings $\tilde\xi_{a}^{\alpha}$ 
and the ``fundamental" couplings $\eta^q$ of the lepton-quark Lagrangian 
(\ref{eff-q}) cannot be established in this simplified model without 
additional assumptions. In particular, this is because of the divergence 
of the quark loops. However, we do not intend to employ this model for 
detailed calculations. Our goal is to illustrate the situation which may 
happen in a true, yet non-existing, theory of hadronization when the 
non-derivative terms in the effective lepton-meson Lagrangian are 
prohibited contrary to the general phenomenological treatment allowing 
such terms in Eq.~(\ref{eff-LV}). 

Let us study the impact of such a situation on $\mu^- - e^-$ 
conversion. To this end we repeat the analysis of the previous section 
for the derivative Lagrangian in Eq.~(\ref{LV-loop}). In the following 
we do not refer to any hadronization model which may be the basis of 
that Lagrangian, treating the couplings ${\tilde\xi}^{\alpha}$ as free 
parameters. We relate them to the \Lfv parameters $\eta$  of the 
quark-lepton Lagrangian in Eq.~(\ref{eff-q}), using the on-mass-shell 
matching condition (\ref{match1}) and relations (\ref{mat-el2}).  
The result is: 
\begin{eqnarray}
{\tilde\xi}_a^{\rho} \, = \, f_{\rho} \,(\eta_{a V}^{u} 
- \eta_{a V}^{d})\,, \,\, 
{\tilde\xi}_a^{\omega} \, = \, 3 \, f_{\omega} \, 
(\eta_{a V}^{u} + \eta_{a V}^{d}) \,, \,\,  
{\tilde\xi}_a^{\phi} \, = \, - 3 \, f_{\phi} \, 
\eta_{a V}^{s} \,. 
\end{eqnarray} 
Note, that the vector meson-quark couplings $g_{_{V qq}}$ introduced 
in Eq.~(\ref{V-q}) do not appear in these formulas explicitly, being 
absorbed, alone with the divergent quark loops, by the couplings 
$f_V$ which are experimentally measurable quantities (\ref{decay}) with 
the values given in Eq.~(\ref{constants}). 

As explained in the previous section the vector meson exchange 
generates the lepton-nucleon Lagrangian (\ref{eff-N}) in second order 
in ${\cal L}_{eff}^{lV} + {\cal L}_{VN}$ shown in Fig.2b. 
The corresponding couplings are given by the expressions:
\begin{eqnarray}\label{alpha-V-ex-2} 
\alpha_{aV[MN]}^{(3)} &=&  - \, 
{\tilde\beta}_{\rho} \, (\eta_{aV}^{u} - \eta_{aV}^{d})\,, \,\,\,  
\alpha_{aV[MN]}^{(0)}  \, = \, - \, 
{\tilde\beta}_{\omega} \, (\eta_{aV}^{u} 
+  \eta_{aV}^{d}) \, - \, {\tilde\beta}_{\phi} \, \eta_{aV}^{s}, 
\end{eqnarray}
where the coefficients ${\tilde\beta}_{_{V}}$ are: 
\ba{beta1}
{\tilde\beta}{_{\rho}} \, = \frac{1}{2}\,  \, \frac{
g_{_{\rho NN}} \, f_{\rho} \, m_{\mu}^2}{{m_{\rho}^2 + m_{\mu}^2}}, \  
{\tilde\beta}{_{\omega}} \, = \, \frac{3}{2} \, \, 
\frac{g_{_{\omega NN}} \, f_{\omega} \,  
m_{\mu}^2}{m_{\omega}^2 + m_{\mu}^2}, \ {\tilde\beta}{_{\phi}} \, = 
\, - \frac{3}{2}  \, \, \frac{g_{_{\phi NN}} \, f_{\phi} \, 
m_{\mu}^2}{m_{\phi}^2 + m_{\mu}^2}\,. 
\ea 
Using the numerical values of the vector meson masses and coupling
constants from Eqs.~(\ref{constants}) and (\ref{VN-couplings}) we obtain 
\begin{eqnarray}\label{beta-til}
&&{\tilde\beta}{_{\rho}} \ \ \ = 7.5\times 10^{-3} \,, \, \, \, \ \ 
{\tilde\beta}{_{\omega}}\ \ \, = 6.6\times 10^{-3} \,,\\ \nn
&& {\tilde\beta}{_{\phi}}^{(1)} = 2.2\times 10^{-3}, \ \ \ \ 
{\tilde\beta}{_{\phi}}^{(2)} = 2.9\times 10^{-5}.
\end{eqnarray} 
Naturally, in the case of the double derivative lepton-meson 
Lagrangian (\ref{LV-loop}) the \Lfv couplings $\alpha$ of the 
corresponding lepton-nucleon Lagrangian (\ref{eff-N}) are much 
smaller, by a factor ${\tilde\beta}_V/\beta_V \, = \, 
(m_\mu/m_V)^2\, \sim \, 10^{-2}$, than in the previously analyzed 
non-derivative case (\ref{alpha-V-ex}). This demonstrates that
the hadronization prescription may have a strong impact on the \Lfv 
new physics contribution to $\mu^- - e^-$ conversion.  

\section{Constraints on \Lfv parameters from muon-electron conversion}

Starting from the Lagrangian (\ref{eff-N}) it is straightforward to 
derive the formula for the total $\mu-e$ conversion branching 
ratio~\cite{Chiang:xz}. In the present paper we focus on the coherent 
process, i.e. ground state to ground state transitions, which is the 
dominant channel of $\mu-e$ conversion exhausting, for the majority of 
experimentally interesting nuclei, more than $90\%$ of the total \m  
branching ratio~\cite{Schwieger:dd}. To leading order of the 
non-relativistic reduction the coherent $\mu-e$ conversion branching 
ratio takes the form~\cite{Kosmas:ch,Chiang:xz} 
\begin{equation} 
R_{\mu e^-}^{coh} \ = \  
\frac{{\cal Q}} {2 \pi \Lambda_{LFV}^4} \  \   
\frac{p_e E_e \ ({\cal M}_p + {\cal M}_n)^2 } 
{ \Gamma ({\mu^-\to capture}) } 
\, , 
\label{Rme}
\end{equation} 
where $p_e, E_e$ are 3-momentum and energy of the outgoing electron.  
The nuclear transition matrix elements ${{\cal M}}_{p,n}$ in 
Eq.~(\ref{Rme}), for the case of a ground state to ground state 
\mbox{\m} transition, are defined as 
\begin{equation}
{\cal M}_{p,n} = 4\pi \int j_0(p_e r) \Phi_\mu (r) \rho_{p,n} (r) r^2  
dr \ , \label{V.1}
\end{equation}
where $j_0$ is the zero-order spherical Bessel function. The quantities 
$\rho_{p,n}(r)$ are the spherically symmetric proton (p) and neutron (n) 
nuclear densities normalized to the atomic number $Z$ and neutron number 
$N$ of the nucleus, respectively. $\Phi_{\mu}(r)$ is the space dependent 
part of the muon wave function.
The factor ${\cal Q}$ in Eq.~(\ref{Rme}) has the 
form~\cite{Kosmas:2001mv} 
\begin{eqnarray}
\hspace*{-1cm}
{\cal Q} &=& |\alpha_{VV}^{(0)}+\alpha_{VV}^{(3)}\ \epsilon|^2 +
|\alpha_{AV}^{(0)}+\alpha_{AV}^{(3)} \epsilon|^2 + 
|\alpha_{SS}^{(0)}+\alpha_{SS}^{(3)} \epsilon|^2 + 
|\alpha_{PS}^{(0)} + \alpha_{PS}^{(3)} \epsilon|^2 
\nonumber  \\ 
\hspace*{-1cm} 
&+& 2{\rm Re}\{(\alpha_{VV}^{(0)}+\alpha_{VV}^{(3)} 
\epsilon)(\alpha_{SS}^{(0)}+ \alpha_{SS}^{(3)} \epsilon)^\ast
+  (\alpha_{AV}^{(0)}+\alpha_{AV}^{(3)}\ \epsilon)(\alpha_{PS}^{(0)} + 
\alpha_{PS}^{(3)}\ \epsilon)^\ast\}\,.
\label{Rme.1} 
\end{eqnarray} 
in terms of the parameters of the lepton-nucleon effective 
Lagrangian (\ref{eff-N}) and the nuclear structure factor 
\begin{eqnarray}\label{factor1}
\epsilon = ({\cal M}_p - {\cal M}_n)/({\cal M}_p + {\cal M}_n) \,. 
\end{eqnarray} 

The nuclear matrix elements ${\cal M}_{p,n}$, defined in 
Eq.~(\ref{V.1}), have been numerically calculated in 
Refs.~\cite{Kosmas:2001mv,Faessler:pn} for the nuclear targets 
$^{27}$Al, $^{48}$Ti and $^{197}$Au, using proton densities $\rho_{p}$ 
from Ref.~\cite{DeJager:qc} and neutron 
densities $\rho_{n}$ from Ref.~\cite{Gibbs:fd} whenever possible. 
The muon wave function $\Phi_\mu(r)$ has been obtained by solving the 
Schr\"ondinger equation with the Coulomb potential produced by the 
densities $\rho_{p,n}$, taking into account the vacuum polarization 
corrections~\cite{Faessler:pn}.  
The results for ${\cal M}_{p,n}$ corresponding to the 
nuclei Al, Ti and Au are given in \mbox{Table 1} where
we also show the muon binding energy $\epsilon_b$ and the experimental 
total rates $\Gamma ({\mu^-\to capture})$ of 
the ordinary muon capture reaction~\cite{Suzuki:1987jf}.

As follows from \mbox{Table 1} the parameter $\epsilon$ in  
Eq.~(\ref{Rme.1}) is small $\epsilon\sim 10^{-1}$ and, therefore,  
the contribution of isovector $\alpha^{(3)}\epsilon$ terms can be  
neglected except for a very special domain in the \Lfv parameter  
space where $\alpha^{(0)}\leq \alpha^{(3)}\epsilon$.  
For this reason the role of the isovector $\rho$-meson exchange  
in $\mu^- - e^-$ conversion is expected to be unimportant except for   
this special case. 

With these matrix elements we find, for the combination of the 
dimensionless vector nucleon couplings $\alpha_{aV}^{(0)}$ $(a = A,V)$ 
and the characteristic \Lfv scale $\Lambda_{LFV}$ in the effective 
lepton-nucleon Lagrangian (\ref{eff-N}) the following limit
\begin{equation}\label{alpha-lim}
\alpha_{aV}^{(0)} \left(\frac{1 \mbox{GeV}}{\Lambda_{LFV}}\right)^2 
\leq 1.2 \times 10^{-12} B(Exp).
\end{equation}
Here the factor $B(A)$ depends on the nuclear matrix element 
${\cal M}(A)= {\cal M}_n(A)+ {\cal M}_p(A)$ 
of the target nucleus $A$ used in an experiment setting the upper limit 
$R_{\mu e}^{A}(Exp)$  on the branching ratio of \m conversion 
$R_{\mu e}^{A}\leq R_{\mu e}^{A}(Exp)$. This factor has the form 
\begin{eqnarray}\label{factor-B}
B(A) = \frac{{\cal M}(A)}{{\cal M}(^{48}Ti)} 
\left(\frac{R_{\mu e}^{A}(Exp)}{1.2 \times 10^{-12}} \right)^{1/2}
\end{eqnarray}  
and allows one to translate the limits in Eq. (\ref{alpha-lim}) to the 
limits of a specific experiment. For the running and forthcoming 
experiments discussed in the introduction this factor takes the values 
\begin{eqnarray}\label{B-nv}
&&\mbox{Eq. (2)}: \ B(^{48}Ti) = 1; \ \ \ \ \ \,  \ \ \ \mbox{Eq.(3)}: 
\  B(^{27}Al)= 7.3\times 10^{-3}; \\ \nn
&&\mbox{Eq. (4)}: \ B(^{197}Au)= 0.57; \ \ \ \mbox{Eq.(5)}: 
\ B(^{48}Ti)\sim 10^{-3}. 
\end{eqnarray}
From the limit in Eq.~(\ref{alpha-lim}) one can deduce the individual 
limits on different terms entering in the expressions for the 
coefficients $\alpha_{aV}^{(0)}$ in the direct nucleon and 
meson-exchange mechanisms, assuming that significant cancellations 
(unnatural fine-tuning) between different terms are absent.  
In this way we derive constraints for the $\eta$ parameters of the
quark-lepton Lagrangian~(\ref{eff-q}) for the meson-exchange 
mechanism (MEM). We present these limits in Table~2 only for the case of 
the model independent approach based on the Lagrangian in 
Eq.~(\ref{eff-LV}). For the parameter $|\eta_{aV}^{s}|$ we derive the 
limits for the two different cases of $g_{_{\phi NN}}$ coupling, given 
in Eq.~(\ref{VN-couplings}). In Table~2 we also show for comparison the 
limits corresponding to the direct nucleon mechanism (DNM) derived in 
Ref.~\cite{Kosmas:2001mv}. With the values of the experimental factor 
$B(A)$ given in Eqs.~(\ref{factor-B}) and (\ref{B-nv}) one can 
translate the limits in Table~2 to the case of a particular experiment.

The limits presented in Table~2 show the importance of the vector
meson exchange mechanism to $\mu^- - e^-$ conversion. The MEM leads 
to the new experimental upper bound on the previously unconstrained 
strange quark \Lfv parameter $\eta_{aV}^{s}$. Also, the MEM limits on 
the \Lfv parameters $\eta_{aV}^{u,d}$ are by a factor $\sim 2$ better 
than the DNM limits. 

The limits in Table~2 represent a general outcome of the \m conversion 
experiments for the \Lfv physics. Here we do not show the limits for 
the scalar quark current coefficients $\eta_{PS}^{q}$ of the Lagrangian 
in Eq.~(\ref{eff-q}), which can be found in Ref.~\cite{Kosmas:2001mv}. 
The limits in Table~2 can be easily translated into limits on the 
parameters of specific \Lfv model predicting the \m conversion. This is 
achieved by adjusting the quark level effective Lagrangian of the model 
to the form of Eq.~(\ref{eff-q}) and by identifying the effective 
parameters $\eta^{(q)}$ with expressions in terms of model parameters. 
Then, assuming absence of cancellation between different terms in these 
expressions, one can extract the upper bounds on the model parameters.

\section{Summary} 

We studied nuclear $\mu^--e^-$ conversion in a general framework 
based on an effective Lagrangian without referring to any specific 
realization of the physics beyond the standard model responsible 
for lepton flavor violation. We demonstrated that the vector 
meson-exchange contribution to this process is significant.  
A new issue of the meson-exchange mechanism in comparison to the 
previously studied direct nucleon mechanism is the presence of the 
strange quark vector current contribution induced by the $\phi$ meson. 
This allowed us to extract new limits on the \Lfv lepton-quark 
effective couplings from the existing experimental data. 
To our best knowledge these limits have not yet been discussed 
in the literature.
We also presented a simplified model of hadronization which leads to 
the derivative lepton-meson effective Lagrangian. The model results in 
a suppression, by a factor of $\sim10^{-2}$, of the vector meson 
contribution to $\mu^- - e^-$ conversion in comparison to the general 
phenomenological treatment. This demonstrates the impact of a 
hadronization prescription on the analysis of new physics in processes 
involving hadrons and nuclei. In the literature it is a common point to 
mention the uncertainties which come from the nuclear structure models. 
We stress that the uncertainties arising at the preceding level dealing 
with the hadronization could be comparable or even larger than 
the nuclear uncertainties. 

\vspace*{1cm}

{\bf Acknowledgments}

\noindent 
This work was supported by the DAAD under contract 
415-ALECHILE/ALE-02/21672, by the FONDECYT projects 1030244, 1030355,  
by the DFG under contracts FA67/25-3, 436 SLK 113/8 and GRK683, by the 
State of Baden-W\"{u}rt\-tem\-berg, LFSP "Low Energy Neutrinos", 
by the President grant of Russia 1743 "Scientific Schools",  
by the VEGA Grant agency of the Slovac Republic  
under contract No.~1/0249/03.

\begin{table}
{\bf Table 1.} Transition nuclear matrix elements ${\cal M}_{p,n}$ 
(in $fm^{-3/2}$) of Eq. (\ref{V.1}) and other useful quantities
(see the text).

\vspace*{.4cm} 

\begin{center}
\begin{tabular}{|r|c|c|c|c|c|}
   &  &  &  &  &  \\
Nucleus & $p_e \, (fm^{-1})$ & $\epsilon_b \, (MeV)$ &
$\Gamma_{\mu c} \, ( \times 10^{6} \, s^{-1})$ &
${\cal M}_p$ & ${\cal M}_n$  \\
\hline
   &  &  &  &  &  \\
$^{27}$Al  & 0.531 &  -0.470 &  0.71 & 0.047 & 0.045   \\
   &  &  &  &  &  \\
$^{48}$Ti  & 0.529 &  -1.264 &  2.60 & 0.104 & 0.127   \\
   &  &  &  &  &  \\
$^{197}$Au & 0.485&   -9.938 & 13.07 & 0.395 & 0.516   \\
\end{tabular}
\end{center}
\end{table}

\begin{table}
{\bf Table 2.} Upper bounds on the \Lfv parameters inferred from the 
SINDRUM II data on $^{48}$Ti [Eq.~(\ref{Ti})] 
corresponding to the direct nucleon mechanism (DNM)
and the meson exchange mechanism (MEM). 
The subscript notation is $a = V,A$. The value in square brackets 
refers to the $g_{_{\phi NN}}^{(1)}$ value of $\phi NN$ coupling
presented in Eq.~(\ref{VN-couplings}). The experimental factor $B(A)$ 
is defined in Eqs.~(\ref{factor-B}) and (\ref{B-nv}).

\vspace*{.4cm} 

\begin{center} 
\begin{tabular}{|c|c|c|} 
Parameter & DNM &MEM \\ 
\hline 
&&    \\ 
$|\eta_{aV}^{u,d}|(1 \, \mbox{GeV}/\Lambda_{LFV})^2$ 
& $8.0\times 10^{-13}\ B(A)$&$3.3\times 10^{-13}\ B(A)$\\  
&&    \\ 
$|\eta_{aV}^{s}|(1 \, \mbox{GeV}/\Lambda_{LFV})^2$& no limits
& $4.0\times 10^{-11}\ B(A)$ \,\, $[6.0\times 10^{-13}\ B(A)]$\\ 
&&    \\ 
\end{tabular}
\end{center} 
\end{table}

\newpage 

\begin{figure}

\noindent {\bf Fig. 1:} 
(a) Photonic (long-distance) and (b) non-photonic (short-distance) 
mechanisms to the nuclear $\mu^-- e^-$ conversion. 
\vspace*{.5cm} 

\noindent {\bf Fig.2:}  
Diagrams contributing to the nuclear $\mu^--e^-$ conversion: 
direct nucleon mechanism (a) and 
meson-exchange mechanism (b). 
\vspace*{.5cm}

\noindent {\bf Fig.3:} 
Generation of the \Lfv lepton-meson effective 
Lagrangian in the simplified model of hadronization.

\end{figure}

\begin{figure}

\vspace*{5cm}
\centering{\epsfxsize=15 cm\epsffile{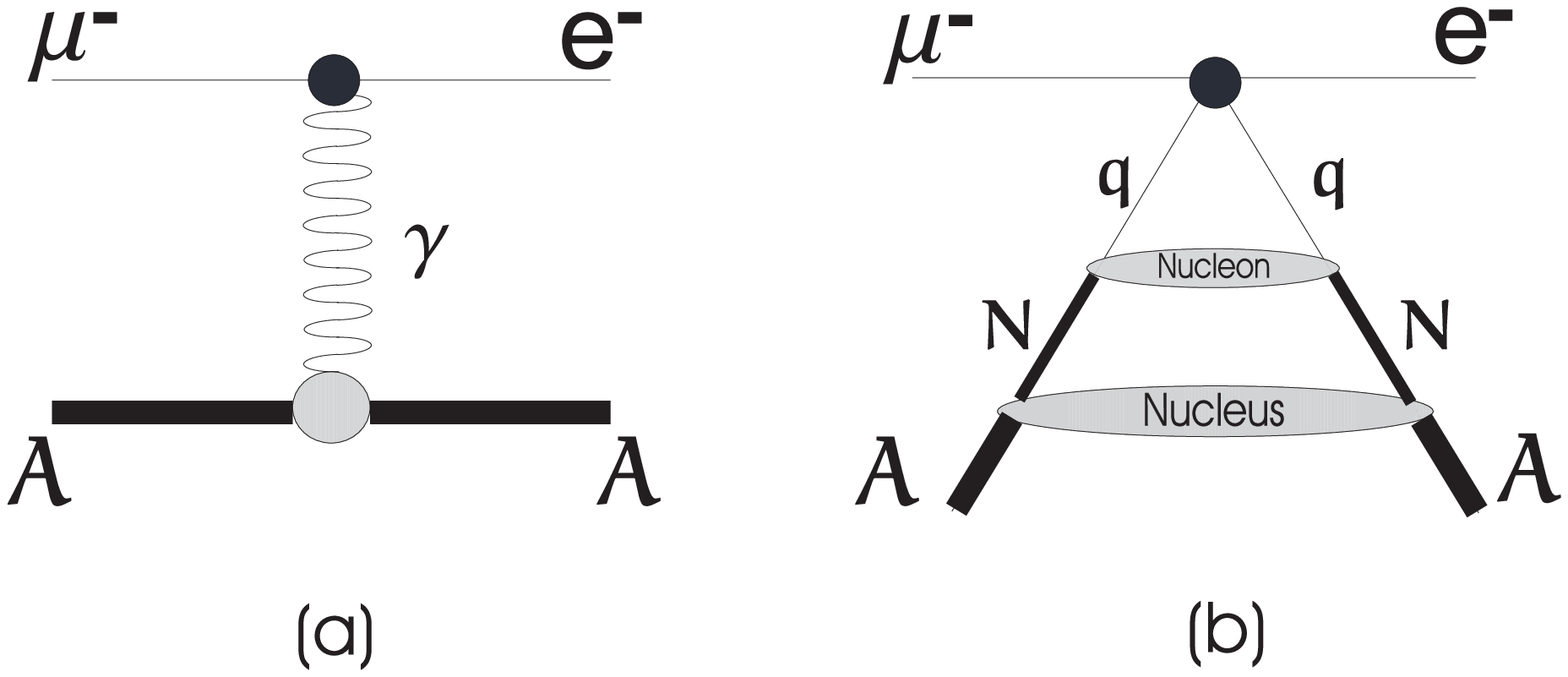}} 
\centerline{\bf Fig.1} 
\end{figure}

\newpage 

\begin{figure}[t]

\vspace*{2cm}

\centering{\epsfxsize=15 cm\epsffile{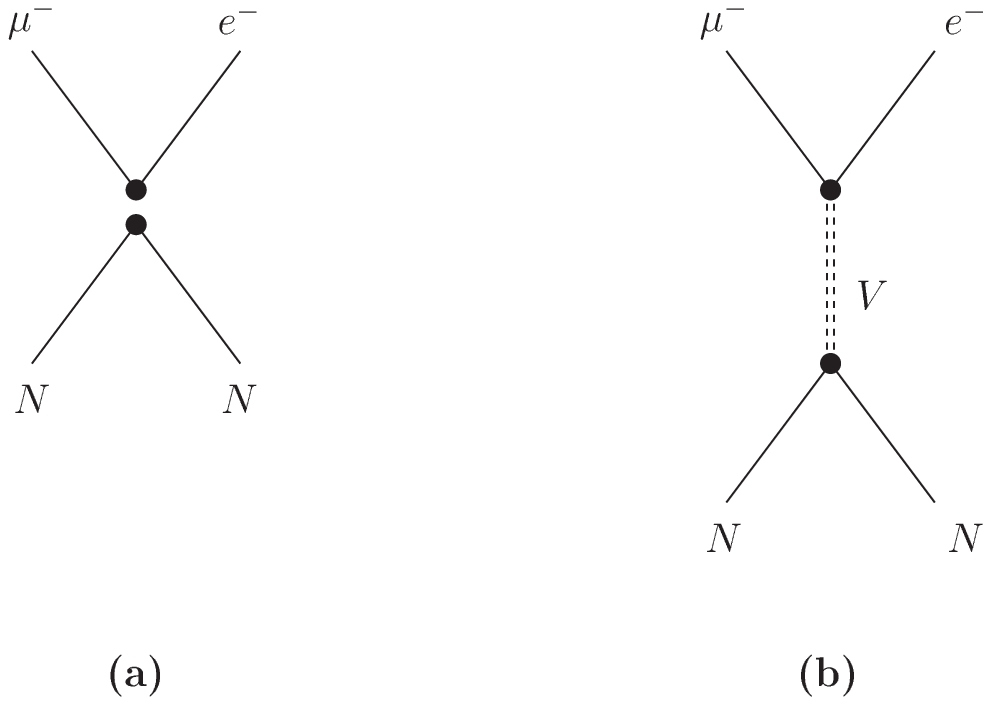}}

\vspace*{.5cm}

\centerline{\bf Fig.2} 
\end{figure}

\newpage

\begin{figure}[t]

\vspace*{16cm}
\hspace*{.5cm}
\centering{\epsfxsize=15 cm\epsffile{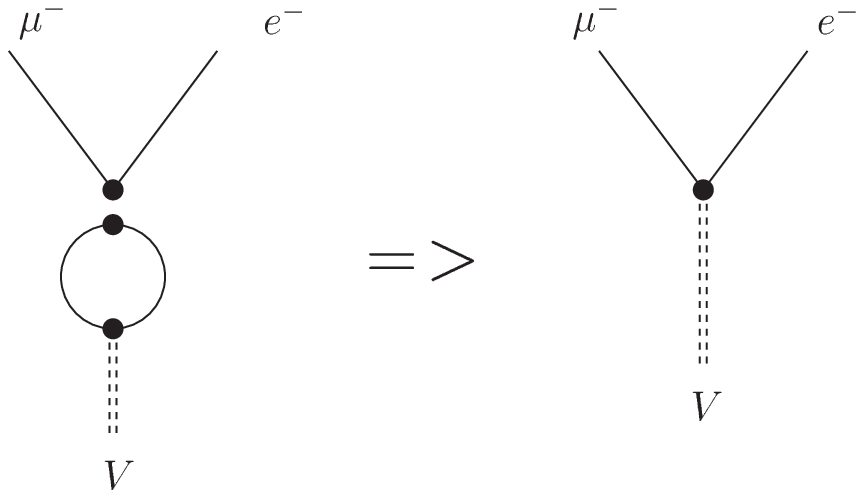}}

\vspace*{-12.5cm}
\centerline{\bf Fig.3} 
\end{figure}


\begin{thebibliography}{99} 
\bibitem{Simkovic:2001fy}
F.~Simkovic, V.~E.~Lyubovitskij, T.~Gutsche, A.~Faessler and S.~Kovalenko,
Phys.\ Lett.\ B {\bf 544}, 121 (2002)
[arXiv:hep-ph/0112277].
%
\bibitem{Kosmas:ch}
T.~S.~Kosmas, G.~K.~Leontaris and J.~D.~Vergados,
Prog. \ Part. \ Nucl. \ Phys. \ {\bf 33}, 397 (1994).  
%
\bibitem{Marciano:conf}
W.J. Marciano, Lepton flavor violation, 
summary and perspectives, Summary talk in the conference on 
"New initiatives in lepton flavor violation and neutrino oscillations 
with very intense muon and neutrino beams", Honolulu-Hawaii, USA, 
Oct. 2-6, 2000 (see also $http://meco.ps.uci.edu/lepton\_workshop.$) 
%
\bibitem{Kuno:1999jp}
Y.~Kuno and Y.~Okada,
Rev.\ Mod.\ Phys.\  {\bf 73}, 151 (2001)
[arXiv:hep-ph/9909265].
%
\bibitem{Honecker:zf}
W.~Honecker {\it et al.} [SINDRUM II Collaboration],
Phys.\ Rev.\ Lett.\  {\bf 76}, 200 (1996); 
A.~van~der Schaaf, Private communication. 
\bibitem{Molzon:sf}
W.~Molzon, 
Nucl.\ Phys.\ Proc.\ Suppl.\  {\bf 111}, 188 (2002).
\bibitem{MECO}J.~Sculli, The MECO experiment, 
Invited talk at~\cite{Marciano:conf}. 
\bibitem{PRIME}Y.~Kuno, The PRISM: Beam-Experiments, 
Invited talk  at~\cite{Marciano:conf}. 
\bibitem{Vintz}P.~Wintz, Status of muon to electron conversion at PSI, 
Invited talk  at~\cite{Marciano:conf}. 
%
\bibitem{Faessler:2004jt}
A.~Faessler, T.~Gutsche, S.~Kovalenko, V.~E.~Lyubovitskij, 
I.~Schmidt and F.~Simkovic,
Phys.\ Lett.\ B {\bf 590}, 57 (2004) 
[arXiv:hep-ph/0403033].
%
\bibitem{Kosmas:2001mv}
T.~S.~Kosmas, S.~Kovalenko and I.~Schmidt,
Phys.\ Lett.\ B {\bf 511}, 203 (2001) 
[arXiv:hep-ph/0102101]; 
Phys.\ Lett.\ B {\bf 519}, 78 (2001).
[arXiv:hep-ph/0107292]. 
%
\bibitem{Faessler:1996ph}
A.~Faessler, S.~Kovalenko, F.~Simkovic and J.~Schwieger,
Phys.\ Rev.\ Lett.\  {\bf 78}, 183 (1997)
[arXiv:hep-ph/9612357].
%
\bibitem{Hagiwara:fs}
K.~Hagiwara {\it et al.}  [Particle Data Group Collaboration],
Phys.\ Rev.\ D {\bf 66}, 010001 (2002).
%
\bibitem{Weinberg:de}
S.~Weinberg,
Phys.\ Rev.\  {\bf 166}, 1568 (1968); 
J.~J.~Sakurai, {\it Currents and Mesons}, 
Chicago Lectures in Physics (The University of Chicago Press,  
Chicago and London, New York, 1967). 
%
\bibitem{Mergell:1995bf}
P.~Mergell, U.~G.~Meissner and D.~Drechsel,
Nucl.\ Phys.\ A {\bf 596}, 367 (1996)
[arXiv:hep-ph/9506375].
%
\bibitem{Kubis:2000zd}
B.~Kubis and U.~G.~Meissner,
Nucl.\ Phys.\ A {\bf 679}, 698 (2001)
[arXiv:hep-ph/0007056].
%
\bibitem{Meissner:1997qt}
U.~G.~Meissner, V.~Mull, J.~Speth and J.~W.~van Orden,
Phys.\ Lett.\ B {\bf 408}, 381 (1997)
[arXiv:hep-ph/9701296].
%
\bibitem{deSwart:gc}
J.~J.~de Swart,
Rev.\ Mod.\ Phys.\  {\bf 35}, 916 (1963); 
H.~Genz and G.~Hohler,
Phys.\ Lett.\ B {\bf 61}, 389 (1976).
%
\bibitem{Gasser:1983yg}
J.~Gasser and H.~Leutwyler,
Annals Phys.\  {\bf 158}, 142 (1984); 
G.~Ecker, J.~Gasser, A.~Pich and E.~de Rafael,
Nucl.\ Phys.\ B {\bf 321}, 311 (1989); 
G.~Ecker, J.~Gasser, H.~Leutwyler, A.~Pich and E.~de Rafael,
Phys.\ Lett.\ B {\bf 223}, 425 (1989);
M.~C.~Birse,
Z.\ Phys.\ A {\bf 355}, 231 (1996) [arXiv:hep-ph/9603251].  
%
\bibitem{Chiang:xz}
H.~C.~Chiang, E.~Oset, T.~S.~Kosmas, A.~Faessler and J.~D.~Vergados,
Nucl.\ Phys.\ A {\bf 559}, 526 (1993).
%
\bibitem{Schwieger:dd}
J.~Schwieger, T.~S.~Kosmas and A.~Faessler,
Phys.\ Lett.\ B {\bf 443}, 7 (1998); 
T.~Kosmas, Z.~Ren and A.~Faessler,
Nucl.\ Phys.\ A {\bf 665}, 183 (2000).
%
\bibitem{Faessler:pn}
A.~Faessler, T.~S.~Kosmas, S.~Kovalenko and J.~D.~Vergados,
Nucl.\ Phys.\ B {\bf 587}, 25 (2000). 
%
\bibitem{DeJager:qc}
C.~W.~De Jager, H.~De Vries and C.~De Vries,
Atom.\ Data Nucl.\ Data Tabl.\  {\bf 36}, 495 (1987).
%
\bibitem{Gibbs:fd}
W.~R.~Gibbs and B.~F.~Gibson,
Ann.\ Rev.\ Nucl.\ Part.\ Sci.\  {\bf 37}, 411 (1987).
%
\bibitem{Suzuki:1987jf} 
T.~Suzuki, D.~F.~Measday and J.~P.~Roalsvig,
Phys.\ Rev.\ C {\bf 35}, 2212 (1987). 
%
\end{thebibliography}
\end{document}